\documentclass[runningheads]{llncs}
\usepackage[T1]{fontenc}
\usepackage{graphicx}
\usepackage{xcolor}
\usepackage{fancyvrb}
\usepackage{subfig}
\usepackage{pgfplots}
\usepackage{booktabs} 
\usepackage{soul}
\PassOptionsToPackage{hyphens}{url}\usepackage{hyperref}
\definecolor{LimeGreen}{rgb}{0.2, 0.8, 0.2}
\definecolor{ProcessBlue}{rgb}{0.0, 0.72, 0.92}

\usepackage{amsmath}
\begin{document}
\title{Movie Recommendation using Web Crawling}
\titlerunning{Movie Recommendation using Web Crawling}
%

\author{Pronit Raj \and
Chandrashekhar Kumar \and
Harshit Shekhar \and\\
Amit Kumar \and
Kritibas Paul \and
Debasish Jana}
\authorrunning{P. Raj, C. Kumar, H. Shekhar, A. Kumar, K, Paul and D. Jana}
%
\institute{
  Department of Computer Science and Engineering, \\Heritage Institute of Technology, Kolkata\\
  \email{\{pronit.raj.cse25, chandrashekhar.kumar.cse25, harshit.shekhar.cse25, amit.kumar.cse25, kritibas.paul.cse25\}@heritageit.edu.in,} \\
  \email{debasish.jana@heritageit.edu}
} 
\maketitle              

\begin{abstract}
In today’s digital world, streaming platforms offer a vast array of movies, making it hard for users to find content matching their preferences. This paper explores integrating real-time data from popular movie websites using advanced HTML scraping techniques and APIs. It also incorporates a recommendation system trained on a static Kaggle dataset, enhancing the relevance and freshness of suggestions. By combining content-based filtering, collaborative filtering, and a hybrid model, we create a system that utilizes both historical and real-time data for more personalized suggestions. Our methodology shows that incorporating dynamic data not only boosts user satisfaction but also aligns recommendations with current viewing trends.

\keywords{Movie Recommendation, Web Crawling, Content-Based and Collaborative Filtering, Machine Learning, Hybrid Model, Data Analysis}
\end{abstract}

\section{Introduction}
\label{sec:intro}
The rise of digital streaming platforms has revolutionized how people access movies and shows, with services from platforms like Netflix, Amazon Prime, Hulu, and others offering extensive catalogs of content~\cite{singh2024impact}. Users increasingly face the challenge of finding the right content among thousands of options, leading to a growing demand for sophisticated recommendation systems~\cite{zangerle2022evaluating}.

Many streaming services are leveraging big data and analytics to shape their strategic decisions. By utilizing vast amounts of user data and advanced algorithms, these platforms can better understand viewer preferences, guiding decisions on content acquisition, investment, and marketing. This approach enables them to provide personalized recommendations that keep existing subscribers engaged while attracting new users in diverse global markets~\cite{shattuc2020netflix}. However, balancing subscriber satisfaction with growth is an ongoing challenge. As users’ tastes evolve and new competitors emerge, recommendation systems must adapt to accurately reflect the preferences of a global, ever-expanding customer base.

The goal is not only to prevent users from leaving but also to keep them engaged and satisfied in an increasingly competitive streaming landscape. 
 
Recommendation engines~\cite{barwal2023impact} are crucial, leveraging data-driven insights and machine learning to curate content that aligns with individual tastes while addressing the diverse demands of global markets.
Traditional recommendation systems fall into three main categories~\cite{javed2021review}: (a) \textbf{Content-based systems}~\cite{maidel2010ontological} recommend items similar to those a user has liked in the past. (b) \textbf{Collaborative systems}~\cite{degemmis2007content} leverage the preferences of other users to make recommendations. (c) \textbf{Hybrid systems}~\cite{zhuhadar2010hybrid} combine both approaches for improved accuracy. However, these systems typically rely on static datasets that fail to capture real-time trends. Our research explores integrating web crawling techniques to collect up-to-date data on movie releases and user preferences, addressing this gap.

\vspace{1mm}
\noindent\textbf{Objective.} The primary objectives of this paper are: 
    \begin{enumerate}
        \item To design a movie recommendation system that combines static and real-time data sources.
        \item To build and evaluate various recommendation techniques, including content-based, collaborative, and hybrid models.
        \item To demonstrate the effectiveness of web crawling in updating movie recommendations to match users' current interests.
    \end{enumerate}

\vspace{1mm}
\noindent\textbf{Scope and Contributions.} This research contributes to the field of movie recommendation systems by integrating static datasets with real-time web-scraped data, offering a more dynamic and responsive solution to user preferences. The paper demonstrates the effectiveness of combining context-based, collaborative, and hybrid models to enhance recommendation accuracy. Additionally, it highlights the importance of web crawling techniques in continuously updating the dataset, ensuring that recommendations reflect the latest trends and user interests. The methodology presented serves as a framework for future research in real-time recommendation systems across various domains.

\vspace{1mm}
\noindent\textbf{Organization of the Paper.} 
The organization of this paper is as follows. Section~\ref{sec:intro} discusses the significance of recommendation systems in the context of digital streaming platforms, highlighting existing challenges and the necessity for real-time data integration to enhance user satisfaction. Section~\ref{sec:relatedwork} the background and development of recommendation systems, covering foundational approaches like content-based and collaborative filtering, along with the emergence of hybrid models to enhance personalization and accuracy. Additionally, it discusses the role of web crawling for integrating real-time data into recommendation systems, comparing existing systems with the proposed approach, which combines both static and dynamic data sources to deliver relevant, up-to-date recommendations. Section~\ref{sec:methodology} provides an in-depth analysis of the challenges associated with traditional movie recommendation systems and the proposed solution for integrating web-scraped data with a static dataset to address these limitations. It outlines the problem statement, the methodologies used for data collection and processing, and the development of a recommendation model that leverages content-based, collaborative, and hybrid filtering approaches. Additionally, Section~\ref{sec:methodology} delves into the technical processes, including data sourcing, scraping techniques, and the structuring of recommendation algorithms, designed to enhance the relevance of movie recommendations. Section~\ref{sec:concl} concludes the paper by discussing the effectiveness of integrating real-time web-crawled data with static datasets for a more dynamic recommendation system. It also outlines future work, including expanding data sources, incorporating advanced modeling techniques, and integrating real-time feedback loops to enhance personalization and adaptability in recommendation systems.
 
\section{Related Work}
\label{sec:relatedwork}
\subsection{Recommendation Systems}
The evolution of recommendation systems is vast, with early systems primarily using content-based approaches, recommending items with similar attributes like genre or keywords~\cite{permana2023movie}. Hooda et al.~\cite{hooda2014study} discussed the adoption of recommender systems in social networks, covering key concepts like collaborative recommendation, content-based recommendation, and hybrid recommendation. Collaborative filtering, however, gained popularity by leveraging user-item interactions to suggest items based on what similar users enjoyed~\cite{wang2023intelligent}. More recently, hybrid models have been developed to combine the strengths of both methods, improving accuracy and personalization. For instance, in movie recommendation, content-based filtering might recommend movies by genre or cast, while collaborative filtering suggests movies based on the viewing patterns of similar users\cite{amangeldieva2024hybrid}.

\subsection{Web Crawling for Real-Time Data}
Web crawling~\cite{olston2010web} has emerged as a crucial method for extracting real-time data from the internet. With improved techniques in data scraping, real-time data can be collected efficiently and used in recommendation systems to capture trends, popularity metrics, and new releases. Khder~\cite{khder2021web} emphasized that code reuse and maintenance are essential in web scraping, as reusable code allows efficient handling of common tasks like site access, while maintenance ensures scrapers remain functional despite changes in site structure or behavior. Recent studies on e-commerce recommendation systems show how web scraping enhances user experiences by integrating frequently updated data on products, which provides relevance and keeps recommendations fresh. Onyenwe et al.~\cite{onyenwe2021developing} applied web crawling and scraping techniques to an e-commerce site to extract HTML data for identifying real-time product updates. Rathod et al.~\cite{rathod2018recommendation} built one system with Python on PyCharm, featureed web scraping and data mining modules to search for user-desired products within specified discounts and budgets, storing preferences for future recommendations and ranking products through a Product Rank Algorithm. Kumar et al.~\cite{kumar2023technique} applied Web scraping useing open-source tools to automate data extraction from the internet, and Python's extensive libraries~\cite{jarmul2017python} to make it a popular choice for web crawling process. Tools like BeautifulSoup~\cite{pant2024web} and Scrapy are commonly used for web crawling~\cite{dikilitacs2023performance}, allowing seamless extraction and processing of structured information.

\subsection{Comparison with Existing Work}
Most current movie recommendation systems rely on static datasets, which limits their ability to provide timely recommendations~\cite{marcuzzo2022recommendation}. Some hybrid models do exist, but they lack real-time updates~\cite{sunny2017implementation}, which can be a drawback for users seeking newly released movies or trending films. This research introduces a novel approach by integrating static and dynamic data, leveraging web-crawled real-time information along with a dataset of movie metadata from Kaggle, to create a more comprehensive and up-to-date recommendation system.

\section{Problem Statement and Proposed Methodology}
\label{sec:methodology}

\vspace{1mm}
\noindent\textbf{Problem Statement.} Existing movie recommendation systems often lack real-time data, leading to a gap in capturing new releases or trending movies, which are crucial for accurate user recommendations. This paper aims to solve the problem by integrating static datasets and web-scraped data to deliver a dynamic recommendation system. By utilizing both historical and fresh data, the aim is to create a system that better reflects current user interests and preferences.
In this paper, we outline a structured approach for building a movie recommendation system through web crawling techniques. It comprises the following stages: dataset creation, data processing, recommendation modeling (encompassing context-based, collaborative, and hybrid approaches).

\subsection{Web Scrapping}
\subsubsection{Data Source} \hfill \break

\noindent To create a reliable and informative dataset, we collected data from Rotten Tomatoes (\url{https://www.rottentomatoes.com/}) and IMDb (\url{https://www.imdb.com/}). These platforms were chosen because of their large repository of movie-related information, including user ratings, reviews, genre classifications, cast, and release dates, which are essential for recommendation modeling. We used web scraping to extract this data programmatically, using Python’s BeautifulSoup~\cite{richardson2007beautiful} and \texttt{requests}~\cite{2024requests} libraries.

Rotten Tomatoes, like many other websites, actively blocks scraping attempts through JavaScript or other mechanisms designed to prevent automated access. As such, prior permission is required to engage in such activities. In the following sections, we provide examples for clarity. The best practice is to use the API access offered by many review and ratings aggregators today to fetch data legally. We present code examples based on a hypothetical or similar site-scraping approach to gather reviews and explain the general steps, keeping ethics in mind.

\vspace{4mm}
\subsubsection{Data Collection Process}\hfill \break

\noindent We scrape the Rotten Tomatoes website to gather key information for each movie, including the title of the movie, the release date, the genre, the ratings of the critics, the ratings of the audience and the top reviews. Additional data, such as cast and crew information, was also collected from IMDb to expand the contextual information on each movie, thus enriching the dataset for more accurate recommendations.

For example, consider the recent release of the movie Oppenheimer. By scraping real-time data from Rotten Tomatoes and IMDb, the system dynamically identified Oppenheimer as trending and updated its recommendations to include this movie for users interested in biographical dramas.

\vspace{1mm}
\noindent\textbf{Rotten Tomatoes.}
Rotten Tomatoes is a widely used website for movie ratings and reviews, providing extensive information on films, TV shows, and audience opinions. It serves as a valuable resource for movie enthusiasts, analysts, and developers interested in trends and viewer preferences. Users can access professional critic reviews and audience scores that reflect public sentiment about films. The available key information includes movie titles, genres, release dates, synopses, and lists of cast and crew. By tracking ratings over time, users can observe trends in genre, director, or actor popularity. The \textit{Tomatometer} displays critic consensus, while audience scores provide information on general public preferences. Data from Rotten Tomatoes can be utilized to build personal databases, compare films across different years, and analyze changes in ratings and genres.

The Rotten Tomatoes API provides \texttt{JSON} feeds for data extraction. We employ the \texttt{requests} and \texttt{simplejson} libraries (packages) to fetch and process data, creating a script to retrieve information on movies that are currently playing. The purpose of this web scraping from the Rotten Tomatoes site is to extract valuable data on movies, such as titles, ratings, and audience opinions, which can help analyze trends and preferences in the film industry. Using the Rotten Tomatoes API, we gather up-to-date information on movies currently playing for further downstream analysis. To scrape information from Rotten Tomatoes, first, we obtain an API key and construct the appropriate API URL to request data on movies in theaters. Next, we use the \texttt{requests} library to make a GET request to the API and retrieve the response content. Then, we parse the \texttt{JSON} response using a library like \texttt{simplejson} to access the relevant movie data, such as titles, and finally loop through the list of movies to print or further process the desired information. A code snippet is given below to show use of APIs to extract needed information. 
{\small
\begin{verbatim}
api_key = "ACTUAL_API_KEY_FROM_ROTTEN_TOMATOES"
api_url = "http://api.rottentomatoes.com/api/public/v1.0/lists/"+
          "movies/in_theaters.json?apikey=%s"
# Make a GET request to the Rotten Tomatoes API
response = requests.get(api_url % api_key)
# Parse the response content to JSON
response_content = response.content
parsed_data = simplejson.loads(response_content)
movies_list = parsed_data["movies"]
# Loop through each movie and print its title
for movie in movies_list:
    print(movie["title"])
\end{verbatim}
}

\noindent We have all the movie information now and thus, we can use the following Python function \texttt{get\_movie\_details()} to extract additional movie details and pass the key and the movie title: The purpose of this code is to retrieve detailed information about a specific movie from the Rotten Tomatoes API using its title and an API key. To perform web scraping, the code first checks for spaces in the movie title and replaces them with plus signs for URL formatting, constructs a request URL with the API key and movie title, sends a GET request to the API, parses the \texttt{JSON} response, and then extracts and displays various details such as the movie's rating, synopsis, cast, runtime, and scores.
{\small
\begin{verbatim}
def get_movie_details(api_key, movie_title):
    if " " in movie_title:
        title_parts = movie_title.split(" ")
        movie_title = "+".join(title_parts)
        
    base_url = "http://api.rottentomatoes.com/api/public/v1.0/movies.json"
    request_url = "%s?apikey=%s&q=%s&page_limit=1"
    request_url = request_url % (base_url, api_key, movie_title)
    
    response = requests.get(request_url)
    json_data = simplejson.loads(response.content)
    
    for film in json_data["movies"]:
        print("Rated: %s" % film["mpaa_rating"])
        print("Movie Synopsis: " + film["synopsis"])
        print("Critics Consensus: " + film["critics_consensus"])
        print("Major Cast:")
        
        for actor in film["abridged_cast"]:
            print("%s as %s" % (actor["name"], actor["characters"][0]))
        
        film_ratings = film["ratings"]
        print("Runtime: %s" % film["runtime"])
        print("Critics Score: %s" % film_ratings["critics_score"])
        print("Audience Score: %s" % film_ratings["audience_score"])
        print("For more information: %s" % film["links"]["alternate"])

\end{verbatim}
}

\begin{figure}[t]
    \centering
    \includegraphics[width=1\columnwidth]{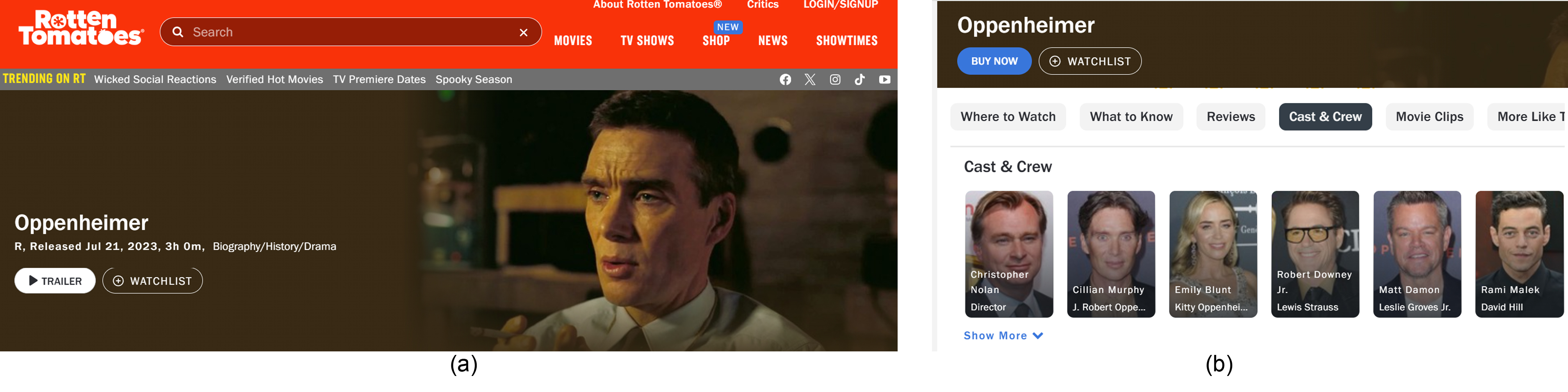} 
    \caption{Snapshot from Rotten Tomatoes site for biopic Oppenheimer(2023) (a) web page (b) Cast \& Crew}
    \label{fig:oppenhiemerOnRottTom}
\end{figure} 

\vspace{-3mm}
\noindent To extract data from Rotten Tomatoes with Python, we use requests to send HTTP requests and use BeautifulSoup to parse the HTML content. At the 96th Academy Awards (2024), Christopher Nolan's epic biopic Oppenheimer, about the father of the atomic bomb, dominated by winning seven Oscars, including Best Picture, Best Director, and Best Actor for Cillian Murphy. The URL for the Rotten Tomatoes page of "Oppenheimer" is: \url{https://www.rottentomatoes.com/m/oppenheimer_2023} (Fig.~\ref{fig:oppenhiemerOnRottTom}). We can send a GET request to the webpage, as below:
\small {
\begin{verbatim}
response = requests.get(url)
soup = BeautifulSoup(response.text, "html.parser")
\end{verbatim}
}
 
\noindent We use the movie's Rotten Tomatoes URL to access the webpage, then parse it to find the element containing the director's name. Note that if Rotten Tomatoes updates its page structure, we might need to adjust the HTML tags and attributes used. We can find the director's name by locating the appropriate HTML element as:
 \small {
\begin{verbatim}
director = soup.find("a", {"data-qa": "movie-info-director"})
\end{verbatim}
}

\noindent Then we can extract and get the director's name as \small{\begin{verbatim}director.get_text() \end{verbatim} }

\noindent\textbf{IMDb.}
IMDb, or the Internet Movie Database, is a wholly owned subsidiary of Amazon.com that serves as a comprehensive resource for information about movies, TV shows, video games, and other forms of entertainment. The site features details such as cast and crew lists, release dates, plot summaries, trailers, trivia, and reviews. The IMDb API allows users to access data related to movies, TV shows, video games, and celebrities. It also provides information on box office performance, cast and crew details, and IMDb ratings.

\begin{figure}[t]
    \centering
    \includegraphics[width=1\columnwidth]{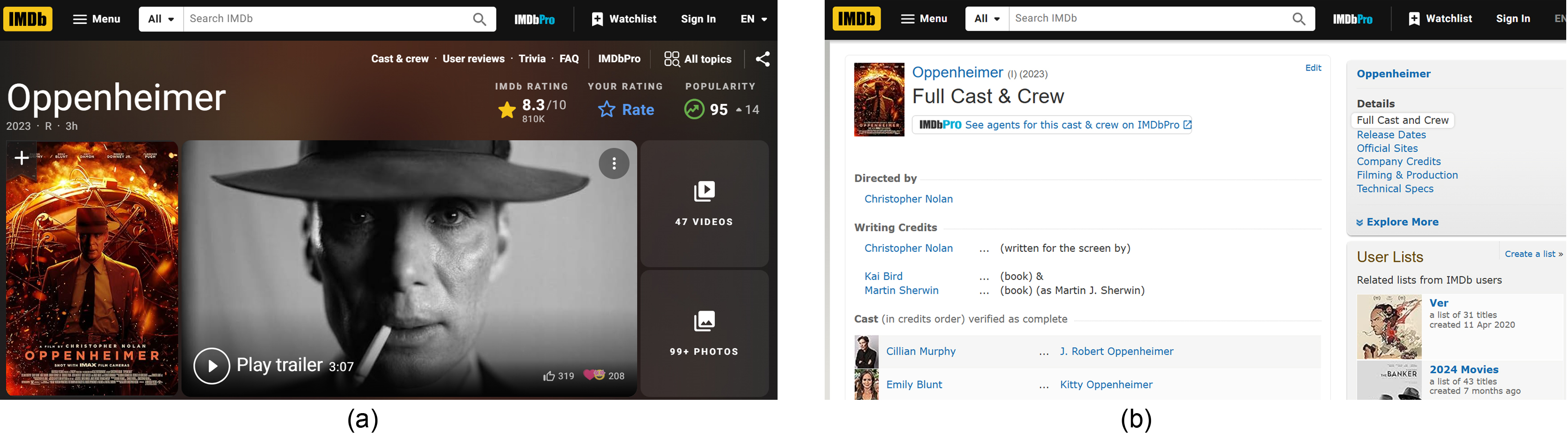} 
    \caption{Snapshot from IMDb site for biopic Oppenheimer(2023) (a) web page (b) Full Cast \& Crew}
    \label{fig:oppenhiemerOnImDb}
    \vspace{-5mm}
\end{figure} 

\small {
\begin{verbatim}
# AWS Data Exchange IMDb endpoint
url = "https://api.imdbws.com/graphql"
# API key (obtained from AWS credentials after subscribing)
headers = {
    "x-api-key": "AWS_API_KEY",
    "Content-Type": "application/json"
}
# GraphQL query to retrieve movie titles
query = """
{
    movies {
        edges {
            node {
                title
                releaseDate
            }
        }
    }
}
"""
# Send the request
response = requests.post(url, json={`query': query}, headers=headers)
# Check if the request was successful
if response.status_code == 200:
    data = response.json()
    # Process the data to extract movie titles
    for movie in data[`data'][`movies'][`edges']:
        print(f"Title: {movie[`node'][`title']}" +
               ", Release Date: {movie[`node']" +
               "[`releaseDate']}")
else:
    print(f"Failed to fetch data: {response.status_code}")
    print(response.text)
\end{verbatim}
}

\noindent This code snippet uses \texttt{requests} package and retrieves movie titles and release dates from the IMDb database using AWS Data Exchange's GraphQL API. It sends a query to the IMDb API endpoint to access specific movie information, displaying the results if successful.

The code starts by setting the IMDb GraphQL API endpoint and preparing authentication headers that include the AWS API key. It then defines a GraphQL query to fetch movie titles and release dates. Using requests.post, it sends an HTTP POST request with the query and headers to the API endpoint. If the request succeeds, the response data is processed and each movie's title and release date are printed; otherwise, an error message is displayed.

\vspace{2mm}
\noindent \textbf{Scrapping.} \texttt{IMDb}'s structure typically doesn’t allow direct scrapping as it’s protected by rules and terms of service. Additionally, \texttt{IMDb} prefers that its data be accessed through its official APIs (available via AWS Data Exchange), which provide a stable and authorized way to access movie data. However, since we wanted to proceed with scraping for research purposes,  here is an example code snippet that demonstrates how we extract the director’s name from the \texttt{IMDb} page for "Oppenheimer" (2023) using BeautifulSoup. Fig.~\ref{fig:oppenhiemerOnImDb} shows snapshot from IMDb site for biopic Oppenheimer(2023) (a) web page (b) Full Cast \& Crew. To run this code, we save the HTML page locally because IMDb blocks bots and scraping scripts. In the HTML of our file, the director's name for Oppenheimer (2023) can be found in a JSON-LD section, specifically in:

\small {
\begin{verbatim}
<script type="application/ld+json">
{
  "@context": "https://schema.org",
  "@type": "Movie",
  "url": "https://www.imdb.com/title/tt15398776/",
  "name": "Oppenheimer",
  ...
  "director": [
    {
      "@type": "Person",
      "url": "https://www.imdb.com/name/nm0634240/",
      "name": "Christopher Nolan"
    }
  ],
  ...
}
</script>
\end{verbatim}
}

\noindent This <script> tag with type="application/ld+json" contains structured data for the page in JSON format. We find the "director" Field: Inside this JSON data, then look for the "director" field, which lists directors as an array of objects. We extract the Name: The director’s name, "Christopher Nolan", is found under the "name" key within the "director" object.
Using BeautifulSoup and Python, we parse this JSON content to extract the director's name. Here’s a Python code snippet that uses json and BeautifulSoup packages to extract the director’s name from the JSON-LD data in our HTML file:

\small {
\begin{verbatim}
# Load the offline saved IMDb HTML page
with open("oppenheimer.html", "r", encoding="utf-8") as file:
    page_content = file.read()
# Parse the HTML content
soup = BeautifulSoup(page_content, "html.parser")
# Find the JSON-LD script tag containing movie information
json_ld_script = soup.find("script", type="application/ld+json")
if json_ld_script:
    # Load JSON data from the script tag
    movie_data = json.loads(json_ld_script.string)
    # Extract the director's name
    director_info = movie_data.get("director", [])
    if director_info:
        director_name = director_info[0].get("name", "Director not found")
        print(f"Director: {director_name}")
    else:
        print("Director information not found")
else:
    print("JSON-LD script with movie data not found")
\end{verbatim}
}

\begin{figure}[t]
    \centering
    \includegraphics[width=1\columnwidth]{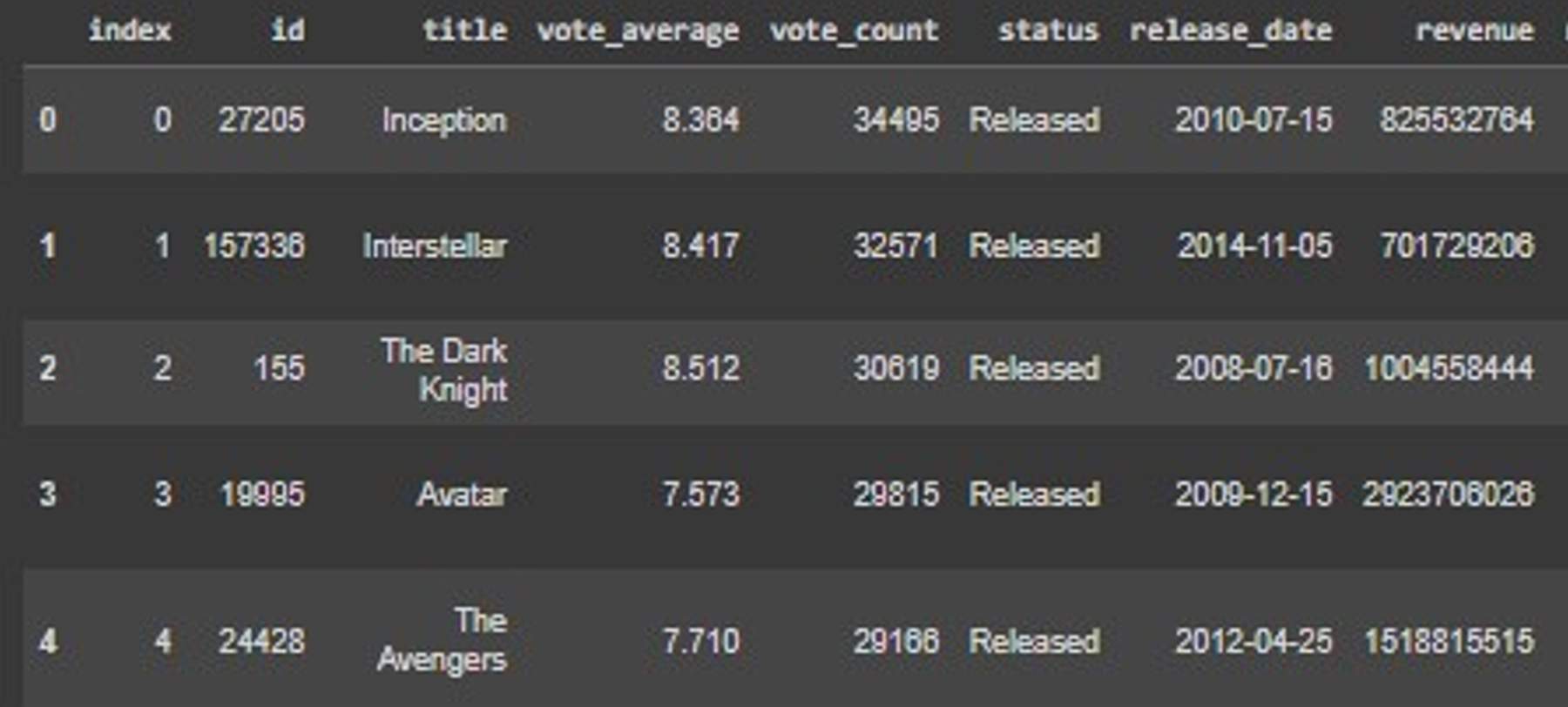} 
    \caption{Top-5 movie recommendations derived from processing the TMDB Movies Dataset 2023}
    \label{fig:snapshot}
    \vspace{-7mm}
\end{figure} 

\subsection{Data Processing through Kaggle}
For developing a movie recommendation system in this paper, we utilize the TMDB Movies Dataset 2023 (tmdb-movies-dataset-2023-930k-movies) from Kaggle, which includes over 930,000 records. This dataset is rich in information, containing attributes like movie titles, genres, release dates, ratings, and cast information, providing a comprehensive base for personalized recommendations. Each record represents a unique movie entry, allowing us to design recommendation algorithms that explore relationships based on both user preferences and the inherent characteristics of movies.
Our approach leverages both content-based filtering and collaborative filtering techniques, along with a potential hybrid model to achieve balanced recommendations. Fig.~\ref{fig:snapshot} shows Top-5 recommendations as obtained from processing the TMDB Movies Dataset 2023.

We further evaluated our approach on the MovieLens 25M dataset, which comprises an extensive collection of 25,000,095 ratings and 1,093,360 tag applications across 62,423 movies. As with Fig.~\ref{fig:snapshot}, which presents our movie recommendation results on TMDB, we generated similar movie recommendations by processing the MovieLens 25M dataset. By evaluating our approach across multiple benchmarks, we demonstrate robustness across various movie pools and feature sets.

\subsection{Recommendation Model} \vspace{1mm}

\noindent\textbf{Content-Based Filtering.}
In content-based filtering, key features from movie attributes (e.g., genres, keywords, descriptions) are extracted to measure similarity among movies using cosine similarity. This approach recommends movies with characteristics similar to those a user has previously enjoyed. Specifically, context-based recommendations suggest movies based on attributes like genre, director, or primary cast members. For example, if a user shows interest in action movies featuring certain actors, the system will recommend other action movies with similar features. Our content-based model utilizes cosine similarity on feature vectors representing genres, directors, and cast members to generate relevant suggestions.

\noindent Example: If a user has expressed interest in films directed by Christopher Nolan, the model prioritizes other movies by Nolan or those with similar stylistic elements. 
When a user searches for science-fiction movies, real-time scraping identifies popular trending movies like Dune: Part Two. By integrating this real-time data with the user's historical preferences, the system provides timely, accurate recommendations.

\vspace{1mm}

\noindent\textbf{Collaborative Filtering.}
Collaborative filtering leverages user interaction data, such as ratings, to recommend based on the assumption that users with similar tastes will enjoy similar content. This technique uses matrix factorization, specifically Singular Value Decomposition (SVD), to break down the user-item interaction matrix and discover latent factors that represent user preferences and movie characteristics. By employing collaborative filtering methods, such as matrix factorization or nearest neighbors, the system recommends movies that align with the viewing behavior of similar users.

\noindent Example: If Users A and B both rate several action-thrillers highly, the system may recommend a top-rated movie from B’s list that A hasn't seen yet.

\vspace{1mm}

\noindent\textbf{Hybrid Model.}
The hybrid model combines content-based and collaborative filtering to provide comprehensive recommendations. Collaborative filtering first generates a recommendation list based on user preferences, which is then refined using specific context-based criteria. This approach combines individual preferences with dataset trends, enhancing recommendations from large datasets like the Kaggle TMDB dataset. By combining content features (such as genres and keywords) and collaborative insights from user interactions, the model effectively adapts to new and recurring user interests.

\noindent Example: For a user who favors high-rated science fiction movies, the hybrid model first identifies top-rated options through collaborative filtering, then refines results with context-based filtering to include positively reviewed films or those by specific directors, creating a more personalized recommendation.

\section{Conclusion and Future Work}
\label{sec:concl}
This research proposes integrating real-time web-crawled data with static datasets to build a dynamic movie recommendation system. By combining content-based, collaborative, and hybrid filtering models, the system provides personalized suggestions based on historical and current trends. Real-time data from sources like Rotten Tomatoes and IMDb ensures the model stays relevant, addressing the limitations of traditional systems. The system could be integrated into platforms like Netflix and Amazon Prime to offer personalized recommendations based on seasonal and holiday-specific trends. Future work could expand data sources, incorporate more streaming platforms, and include user feedback from social media. Additionally, integrating multiple datasets like MovieLens and TMDB with real-time data could further improve recommendation accuracy. Advanced techniques such as neural collaborative filtering and deep learning models, along with real-time feedback loops, would enhance adaptability and enable more sophisticated recommendation systems.

\bibliographystyle{splncs04}
\bibliography{mybib}
\end{document}